\documentclass[twocolumn,floatfix,prl,aps,showpacs]{revtex4-1}
\usepackage{graphicx,epsf,amsmath,amssymb}
\usepackage{bm,bbm}
\usepackage{multirow}
\usepackage{enumerate}
\usepackage{hyperref}
\hypersetup{
    bookmarks=true,         % show bookmarks bar?
    unicode=false,          % non-Latin characters in Acrobat's bookmarks
    pdftoolbar=true,        % show Acrobat's toolbar?
    pdfmenubar=true,        % show Acrobat's menu?
    pdffitwindow=false,     % window fit to page when opened
    pdfstartview={FitH},    % fits the width of the page to the window
    pdftitle={Magnetic-field induced relativistic properties in type-I and type-II Weyl semimetals},    % title
    pdfauthor={S.Tchoumakov, M.Civelli, M.O.Goerbig},     % author
    pdfsubject={Report},   % subject of the document
    pdfcreator={S.Tchoumakov, M.Civelli, M.O.Goerbig},   % creator of the document
    pdfproducer={S.Tchoumakov, M.Civelli, M.O.Goerbig}, % producer of the document
    pdfkeywords= {electrons} {spectroscopy} % list of keywords
    pdfnewwindow=true,      % links in new window
    colorlinks=true,       % false: boxed links; true: colored links
    linkcolor=blue,          % color of internal links
    citecolor=red,        % color of links to bibliography
    filecolor=magenta,      % color of file links
    urlcolor=cyan           % color of external links
}

\usepackage{color}

\newcommand{\omegab}{\mbox{\boldmath $\omega $}}
\newcommand{\bone}{\mathbbm{1}}
\newcommand{\BEDT}{$\alpha$-(BEDT-TTF)$_2$I$_3$}
\newcommand{\Emath}{\mathcal{E}}
\newcommand{\Nmath}{\mathcal{N}}

\newcommand{\bk}{{\bf k}}
\newcommand{\bt}{{\bf t}}
\newcommand{\be}{{\bf e}}

\newcommand{\br}{{\bf r}}
\newcommand{\bA}{{\bf A}}
\newcommand{\bB}{{\bf B}}

\begin{document}

\title{Magnetic-field-induced relativistic properties in type-I and type-II Weyl semimetals} 

\author{Serguei Tchoumakov, Marcello Civelli and Mark O. Goerbig}
\affiliation{Laboratoire de Physique des Solides, CNRS UMR 8502, Univ. Paris-Sud, Universit\'e Paris-Saclay F-91405 Orsay Cedex, France}
\date{\today}

\begin{abstract}
We investigate Weyl semimetals with tilted conical bands in a magnetic field. Even when the cones are overtilted (type-II Weyl semimetal), Landau-level
quantization can be possible as long as the magnetic field is oriented close to the tilt direction. Most saliently, the tilt can be described within the relativistic
framework of Lorentz transformations that give rise to a rich spectrum, displaying new transitions beyond the usual dipolar ones in the optical conductivity. 
We identify particular features in the latter that allow one to distinguish between semimetals of different types. 
\end{abstract}


\maketitle

The discovery of massless ultrarelativistic electrons in graphene \cite{monolayer1,monolayer2} has triggered a tremendous interest in novel types of semimetallic 
phases in condensed-matter systems \cite{WSM0,WSM01,WSM02}. A particularly intriguing class is that of electronic systems with tilted Dirac cones that were first investigated
in the framework of the quasi-twodimensional (2D) organic crystal \BEDT\ \cite{kobayashi}. In contrast to graphene where the Dirac cones are situated at
high-symmetry points \cite{mark2}, tilted Dirac cones in \BEDT\ arise and migrate in wavevector space as a function of pressure applied to the system. 
One of the most remarkable consequences of the tilt is unveiled in the presence of a magnetic field $B$. Indeed, the tilt renormalizes the cyclotron frequency
by the same factor as an inplane electric field $\Emath$ that was investigated in a covariant description by Lukose et al. \cite{lukose}. This hints at
an intimate relation between the tilt of the Dirac cones and relativistic transformations that
was later investigated in the framework of magnetotransport \cite{mark1} and magnetooptics \cite{judit}.

Indeed, one may naturally ask what happens when the Dirac cones are \textit{overtilted} such that the original
reciprocal isoenergy trajectories are no longer closed ellipses but open hyperbolas \cite{mark2}. 
In the presence of a magnetic field, the tilted Dirac cones may be characterized with the help of Lorentz transformations.
As we discuss in more detail below, this can be
achieved if one identifies the tilt with an effective electric field $\Emath$. Below the critical tilt, we are confronted with a ``weak'' electric
field that allows for a Lorentz boost to a frame of reference where $\Emath$ effectively vanishes. In this so-called \textit{magnetic regime} \cite{jackson},
the electronic orbits are then still closed (cyclotron orbits) and their energy is quantized into Landau level (LLs) \cite{goerbigRev}.
In contrast, such a boost is not possible for overtilted
cones that corresponds to a ``strong'' electric field. Via a Lorentz transformation, one can now simply find a frame of reference where the $B$-field vanishes. This
\textit{electric regime} is characterized by open orbits that prevent LL quantization.
Whereas overtilted cones have not been found in 2D materials, following the proposal  by Soluyanov et al. \cite{wsm2}, several 3D systems have been identified during 
the last months as possible candidates for representing such phase \cite{exp1,exp2,exp3,exp4}. Furthermore, these phases, coined type-II Weyl semimetals 
(WSM), have been
classified from a topological point of view \cite{muechler,trivedi}.

In the present paper, we show that the above classification needs to be revisited in 3D systems that present a remarkably rich behavior. Indeed, we find that a type-II 
WSM can undergo a transition from the magnetic to the electric regime as a function of the angle between the magnetic field and the tilt direction. Whereas the electric
regime is at first sight the natural regime of a type-II WSM, the latter can nevertheless show LL quantization--the tilt is then most prominent in 
the direction of the $B$-field, and the original overtilt is well represented by a 
%is then present in the 
onedimensional band in the latter direction.
This situation needs to be contrasted with that of at type-I WSM that
is always in the magnetic regime regardless of the $B$-field direction. However, one
can, at least in principle, induce transitions between WSM of the two different types by the
application of an additional electric field with a nonzero component perpendicular to the $B$-field. Finally, we investigate in detail the magneto-optical 
signatures in the magnetic regime, which allow for a distinction between  type-I and type-II WSM. 
Most saliently, we find, similarly to tilted Dirac cones in 2D materials \cite{judit}, that 
a magnetic field applied in a direction different from the tilt yields magnetooptical transitions beyond the usual dipolar ones that couple LLs with
adjacent indices, $n\rightarrow (n\pm 1)$. 

Tilted Dirac cones in 3D systems can be modeled conveniently by the Hamiltonian
\begin{equation}\label{eq:ham1}
 H= \omegab_0\cdot \mathbf{k} \bone + \sum_{\mu = 1}^3 v_{\mu} k_{\mu} \hat{\sigma}_{\mu},
\end{equation}
in terms of the three Pauli matrices $\hat{\sigma}_{\mu}$ ($\mu = 1,2,3$) and the anisotropic Fermi velocities $v_{\mu}$. Here and in the following,
we use a system of units where $\hbar=1$. Furthermore, we combine the 
tilt velocities $\omegab_0=(\omega_{0x},\omega_{0y},\omega_{0z})$ into the tilt parameter vector 
\begin{equation}\label{eq:tilpar1}
 \mathbf{t} = \left( \frac{\omega_{0x}}{|v_{x}|}, \frac{\omega_{0y}}{|v_{y}|}, \frac{\omega_{0z}}{|v_{z}|}\right),
\end{equation}
which allows us to distinguish a type-I WSM ($|\bt|<1$) from a type-II WSM ($|\bt|>1$), in analogy with the 2D case \cite{mark2}. In order to account for a magnetic
field in the $z$-direction, we use the Peierls substitution $\bk \rightarrow (q_x,q_y,k_z)$, with $q_{x/y}=k_{x/y}+eA_{x/y}(\br)$ and $\bB=\nabla\times \bA(\br)$.
As a consequence of the $\mathbf{r}$-dependence of the vector potential, the $x$- and $y$ components of the new momenta become operators that no longer commute,
$[q_x,q_y]= -i\mathrm{sign} (v_xv_yB)/l_B^2$, in terms of the magnetic length $l_B=1/\sqrt{e|B|}$. In the following, we consider the tilt to have no
component in the $y$-direction--this 
can always be achieved via a rotation and a possible rescaling of the momenta--and the Hamiltonian thus reads (see Supplementary Material)
\begin{equation}\label{eq:ham2}
 H_B=(\omega_{\perp}q_x + \omega_{0z}k_z)\bone + v_\perp (q_x\hat{\sigma}_x + q_y\hat{\sigma}_y) +v_zk_z\hat{\sigma}_z,
\end{equation}
where $v_\perp = \sqrt{|v_xv_y|}$ is the average Fermi velocity in the $xy$-plane, and $\omega_{\perp}=v_\perp\sqrt{(\omega_{0x}/v_x)^2+(\omega_{0y}/v_y)^2}$ 
is a rescaled tilt velocity. 

While the Hamiltonian (\ref{eq:ham2}) can in principle, within a lengthy calculation, be solved by the introduction of the usual ladder operators 
$\hat{a}^{\pm}=l_B(q_x \pm i q_y)/\sqrt{2}$, similarly to the 2D case \cite{perez,molinari}, a more elegant method consists of using a hyperbolic transformation
to change the eigenvalue equation $(H_B - E\bone) |\Psi\rangle =0$ into 
\begin{equation}\label{wsma}
 \left( e^{\frac{\theta}{2} \hat{\sigma}_x}H_B e^{\frac{\theta}{2} \hat{\sigma}_x} - E e^{\theta \hat{\sigma}_x} \right) |\tilde{\Psi}\rangle = 0,
\end{equation}
where $|\tilde{\Psi}\rangle = \Nmath \exp(-\theta\hat{\sigma}_x/2)|\Psi\rangle$, and $\Nmath$ is a normalization constant required since the hyperbolic
transformation does not preserve the norm of the wave functions. This transformation is nothing other than a Lorentz boost in the $x$-direction in terms
of the relativistic parameter $\tanh\theta=\beta$ \cite{lukose}, which reads $\beta = -\omega_{\perp}/v_\perp $ in the magnetic regime ($|\beta|<1$)
and $\beta = -v_\perp/\omega_{\perp}$ in the electric regime ($|\beta|>1$) \cite{jmlv}. As mentioned in the introduction, only the magnetic regime allows for
quantized LLs, and we concentrate henceforth on this regime. In the transformed frame of reference, the eigenvalue equation (\ref{wsma}) 
reads 
\begin{equation}
 \left[ \gamma (\omega_{0z} \tilde{k}_z - E) \bone + v_\perp (\tilde{q}_x\hat{\sigma}_x + \tilde{q}_y\hat{\sigma}_y) +v_z\tilde{k}_z\hat{\sigma}_z \right] 
 |\tilde{\Psi}\rangle = 0,
	\label{mal}
\end{equation}
where $\gamma=(1-\beta^2)^{-1/2}$ is the dilatation factor. 
The tilde indicates transformed wave vectors, $\tilde{q}_x=q_x/\gamma + \gamma\beta(\omega_{0z}k_z -E)/v_\perp$, $\tilde{q}_y=q_y$, and $\tilde{k}_z=k_z$,
in agreement with a Lorentz boost in the $x$-direction \cite{jackson}.

\begin{figure}[thb]
	\centering
	\includegraphics[width=0.334\textwidth]{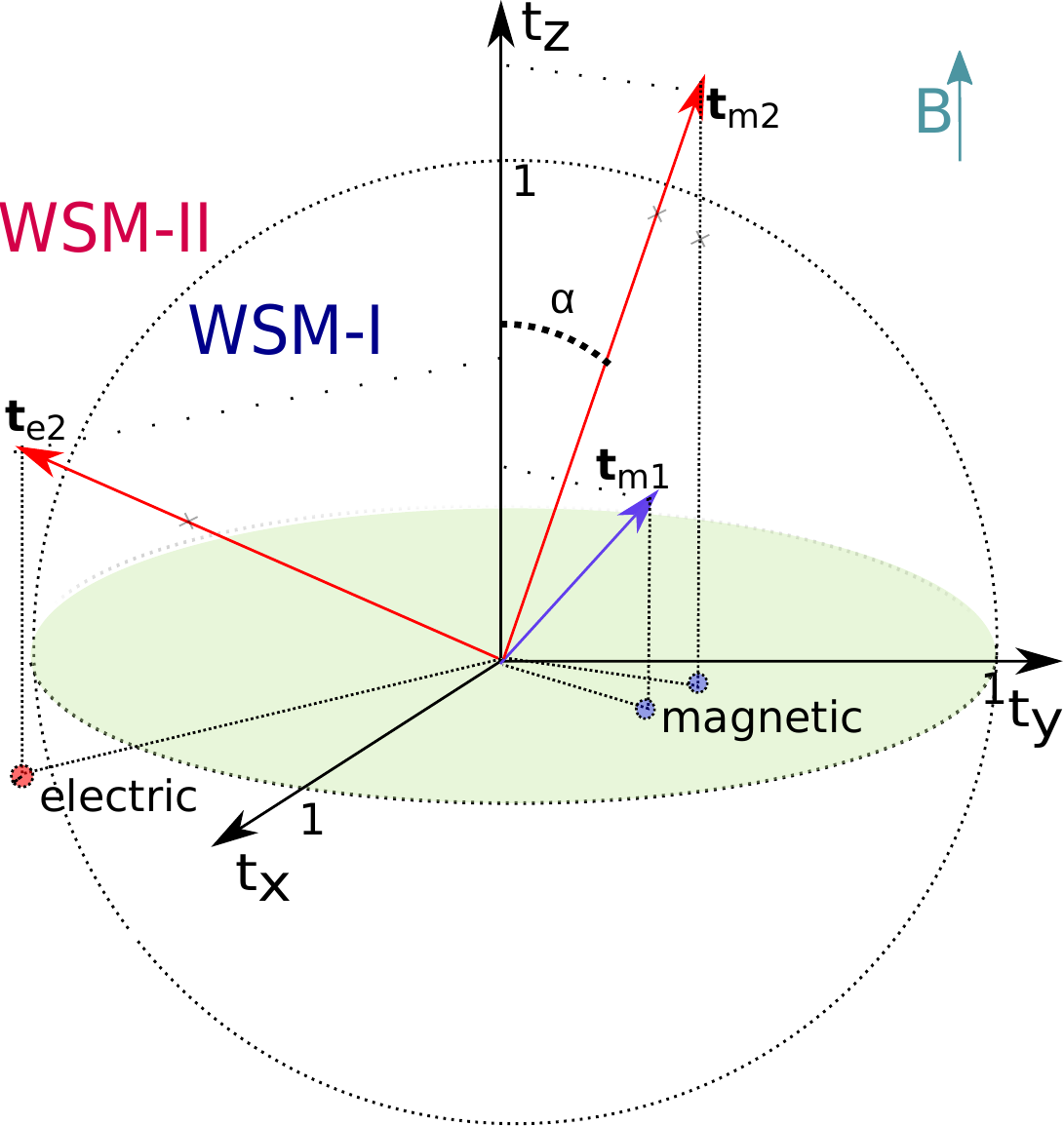}
	\caption{(Color online) Graphical representation of the tilt-parameter vector $\mathbf{t}$.
	The unit sphere $|\mathbf{t}| = 1$ represents the boundary between type-I (blue arrow) and type-II (red arrows) WSM. 
	In the presence of a magnetic field along the $z-$axis, the projection of a tilt-parameter vector on the $xy$ plane identifies the regime, 
	the magnetic regime (\emph{i.e.} inside the green disc) or in the electric regime (outside the disc). The vectors $\bt_{m1}$, $\bt_{m2}$, and
	$\bt_{e2}$ indicate a type-I WSM, a type-II WSM in the magnetic regime, and a type-II WSM in the electric regime, respectively.}
	\label{fig:tilspace}
\end{figure}

One notices that the eigenvalue equation (\ref{mal}) now only contains the noncommutative wavevector components $\tilde{q}_x$ and $\tilde{q}_y$ in the offdiagonal
matrix elements. It can therefore easily be solved by the introduction of the standard ladder operators, but with a renormalized $B$-field because
the commutation relations between the transformed wavevector components read $[\tilde{q}_x,\tilde{q}_y]=-i\mathrm{sign} (v_xv_yB)/\gamma l_B^2$ and
$[\tilde{q}_x,k_z]=[\tilde{q}_y,k_z]=0$. One thus finds the LLs
\begin{eqnarray}
\nonumber\label{eq:LLs}
 E_{n,\lambda}(k_z) &=& \omega_{0z} k_z +\lambda \frac{1}{\gamma} \sqrt{v_z^2k_z^2 + \frac{ 2 eBv_\perp^2}{\gamma}n } ~~~\mathrm{for}~n>0,\\
 E_0(k_z) &=& \left[\omega_{0z} - \mathrm{sign} (v_xv_yB)v_z/\gamma\right]k_z~~~\mathrm{for}~n=0,
\end{eqnarray}
where $\lambda=\pm 1$. 
Let us first discuss the different regimes. As mentioned above, LL quantization is only possible in the magnetic regime. Similarly to the tilt
parameter (\ref{eq:tilpar1}), this can conveniently be described in terms of the inplane tilt vector, 
\begin{equation}\label{eq:tilpar2}
 \bt_\perp = \frac{\bt \times \bB}{B} =\left( \frac{\omega_{0x}}{|v_x|},\frac{\omega_{0y}}{|v_y|}\right).
\end{equation}
Its norm is precisely $|\beta|$ and can be related to the tilt parameter (\ref{eq:tilpar1}) and the angle $\alpha$ between the magnetic
field and the tilt direction, via $|\beta|=|\bt_\perp|=|\bt||\sin\alpha|$. These geometric relations are shown in Fig. \ref{fig:tilspace}, where the 
sphere $|\bt|=1$ indicates the border between a type-I (inside) and a type-II WSM (outside). Similarly, the magnetic regime is represented by the inside (green)
of the circle $t_x^2+t_y^2=1$ while the electric regime is situated outside. Whereas a type-I WSM, represented by the blue vector $\bt_{m1}$
inside the sphere, has always a projection to the $xy$-plane inside the sphere, i.e. in the magnetic regime, one needs to distinguish two situations 
for type-II WSM. Below a critical angle $\alpha_c$, given by
\begin{equation}\label{eq:critang}
 \left|\sin\alpha_c\right|<1/|\bt|,
\end{equation}
the projection of the tilt vector ($\bt_{m2}$ in Fig. \ref{fig:tilspace}) is within the sphere and thus in the magnetic regime. However, if the 
condition (\ref{eq:critang}) is not satisfied, the projection (of $\bt_{e2}$) is outside the unit circle, and the system is thus in the electric regime.

\begin{figure}[thb]
	\centering
	\includegraphics[width=0.22\textwidth]{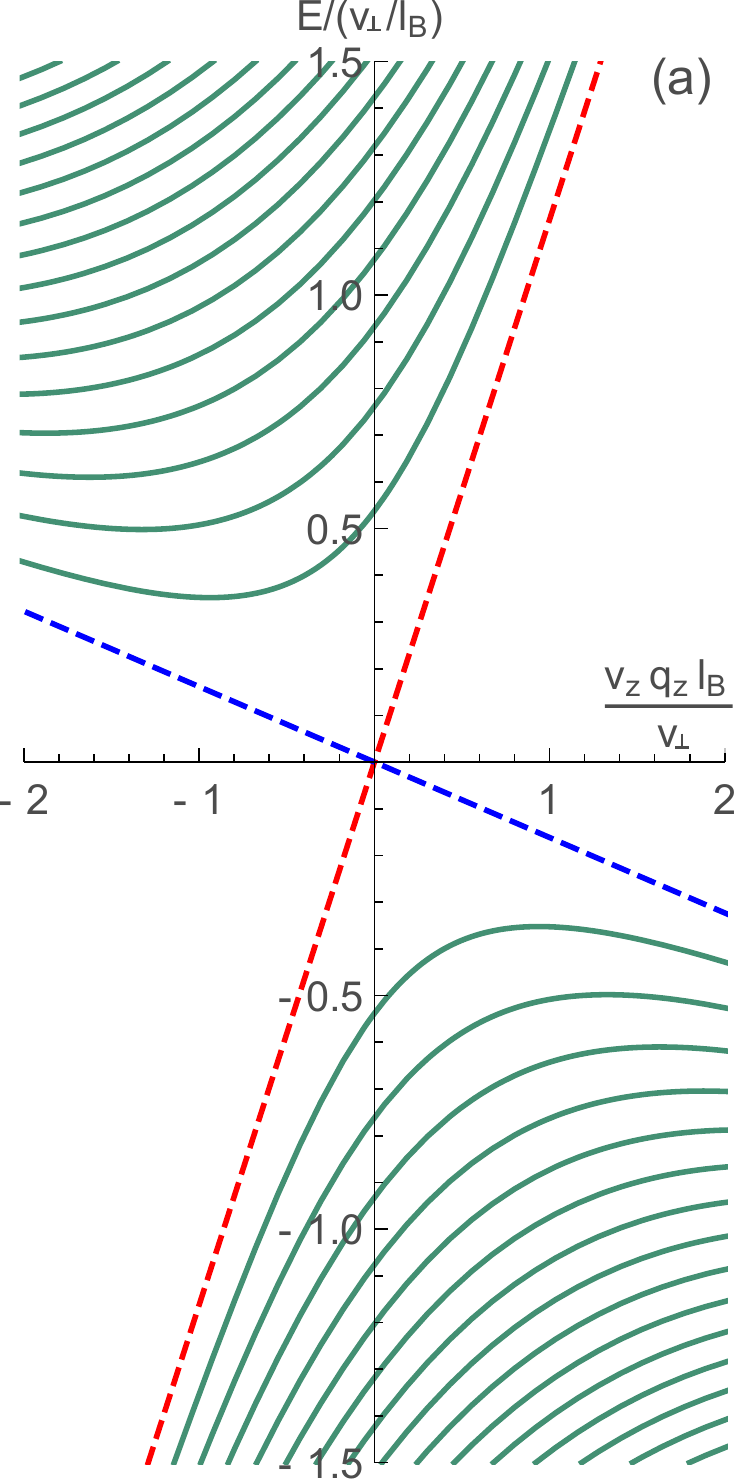}
	\includegraphics[width=0.22\textwidth]{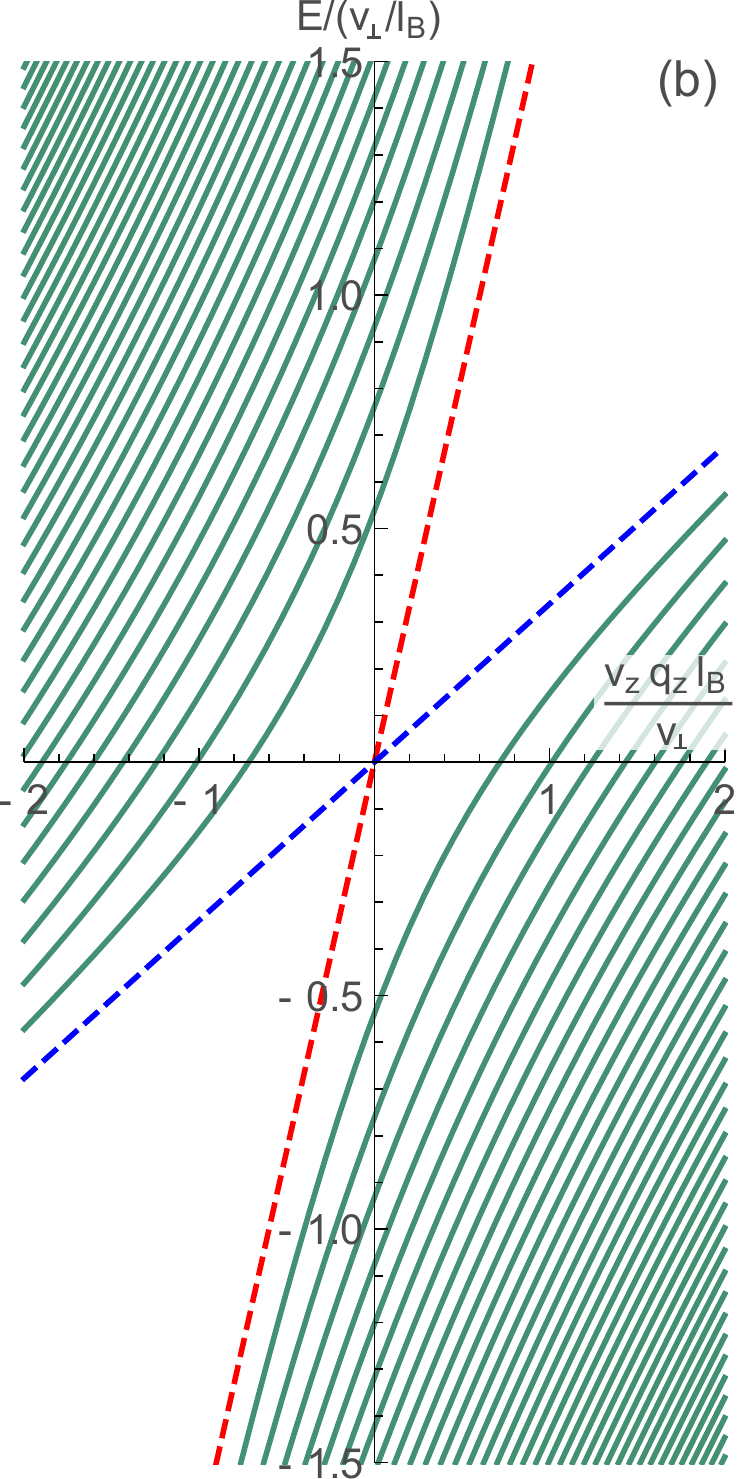}
	\caption{(Color online) LL spectrum of a tilted Weyl cone for $\beta=0.75$, 
	in units of $v_\perp/l_B\simeq 26\, {\rm meV} \times v_\perp[10^6{\rm m/s}]/\sqrt{B[{\rm T}]}$.
	The green lines are $n > 0$ LLs and the blue 
	(red) line represents the $n = 0$ state for $\textrm{sign}(v_xv_yv_zB) = 1$ ($\textrm{sign}(v_xv_yv_zB) = -1$). The left figure 
	corresponds to the case of a type-I Weyl semimetal where the $n = 0$ LL group velocity changes sign with chirality. 
	On the contrary, the right figure is for a type-II Weyl semimetal, where the $n = 0$ 
	slope representing the LL group velocity direction is independent of chirality.}
	\label{figsc}
\end{figure}

The LL spectrum (\ref{eq:LLs}) is represented in Fig. \ref{figsc} for a type-I WSM [panel (a)] and a type-II WSM [panel (b)] in the magnetic regime. One 
retrieves the known 2D results for $k_z=0$ \cite{mark2,molinari,goerbigRev}, i.e. a graphene-like LL spectrum with $E_{n,\pm} \propto \pm (1-\beta^2)^{3/4} \sqrt{Bn}$,
where the spacing is reduced by the relativistic factor $(1-\beta^2)^{3/4}$. In the 3D case, these LLs evolve into 1D bands with a dispersion in $k_z$. It is 
precisely this dispersion that bares information about the underlying WSM type since their bands are also tilted and can be seen as gapless ($n=0$) or
gapped ($n\neq 0$) Dirac cones with a tilt below (type-I) or above (type-II) the critical value. 
Their tilt is characterized by the 1D tilt parameter $t_z=\omega_{0z}/|v_z'|$, in terms of the velocity $v_z'=\sqrt{1-\beta^2}v_z$,
\begin{equation}
 t_z=\frac{t |\cos(\alpha)|}{\sqrt{1 - t^2 \sin(\alpha)^2}}.
\end{equation}
One thus finds that for any angle $\alpha$ one has  $t_z < 1$ for a type-I and $t_z > 1$ for a type-II WSM, i.e. the type of the 1D bands is the same as the 
original 3D WSM in the absence of a magnetic field. 
 
A naturally arising question is whether one can also induce a transition from the magnetic to
the electric regime in a type-I WSM. From a theoretical point of view, this can in principle be achieved by an electric field 
with a nonzero component perpendicular to $\bB$. Let us 
consider for simplicity an electric field $\Emath$ in the $y$-direction that yields a supplementary term $V=e\Emath y\bone$ to Hamiltonian (\ref{eq:ham1}). In the 
Landau gauge $\bA=-By\be_x$, the modified Hamiltonian reads $H_{\Emath,B}=H_B' - (\Emath/B)k_x\bone$, where $H_B'$ has the same structure as Eq. (\ref{eq:ham2}) if
we replace the tilt velocity \cite{mark1} $\omegab_0\rightarrow \omegab=\omegab_0 - \Emath\times \bB/B^2$ or equivalently the tilt parameter 
$\bt\rightarrow \bt_\Emath$, with
\begin{equation}
 \mathbf{t}_\Emath = \mathbf{t} - 
 \frac{1}{B^2}\left( \frac{\left(\mathcal{E}\times\mathbf{B}\right)_x}{|v_{x}|}, \frac{\left(\mathcal{E}\times\mathbf{B}\right)_y}{|v_{y}|}, 
 \frac{\left(\mathcal{E}\times\mathbf{B}\right)_z}{|v_{z}|}\right).
\end{equation}
A type-I WSM can therefore undergo a transition from the magnetic to the electric regime for a judicious choice of the electric field if the 
criterion (\ref{eq:critang}) is satisfied in terms of the tilt parameter $\bt_\Emath$, inducing a breakdown of the LL spectrum \cite{lukose,mark1}. 
From an experimental point of view, however, the situation
is more involved because of the screening of the electric field due to free carriers both in the bulk and at the surfaces. 
Further studies are thus required to establish the feasibility of such experiments.

\begin{figure}[tb]
	\centering
	\includegraphics[width = 0.45\textwidth]{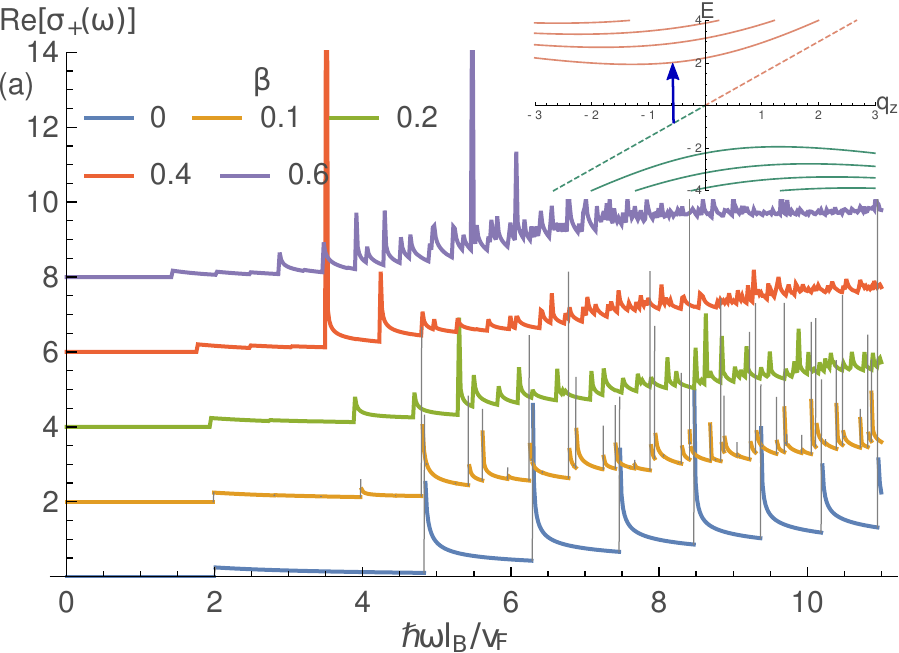}
	\includegraphics[width = 0.45\textwidth]{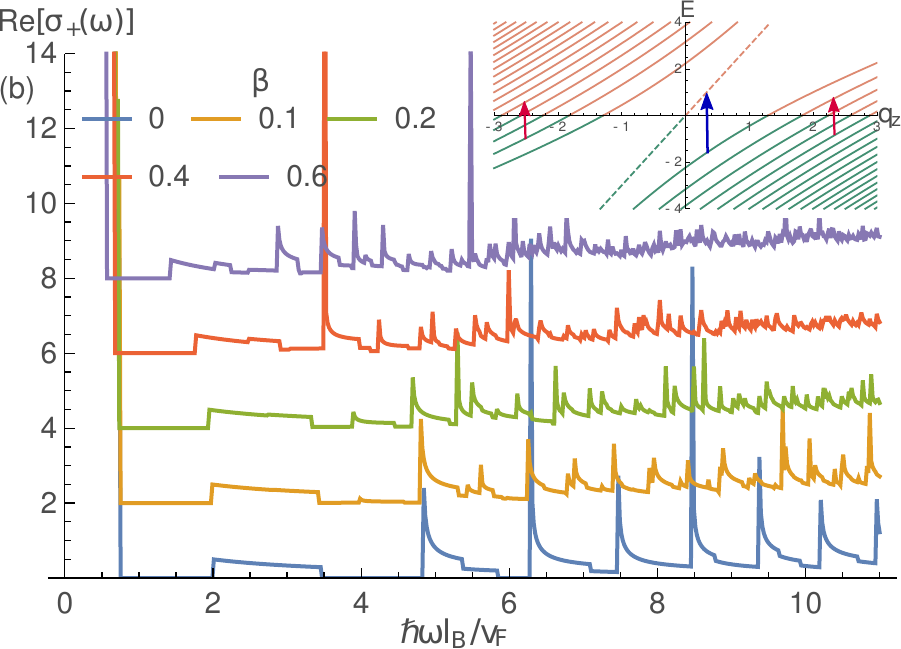}
	\caption{(Color online) Optical conductivity for the $\mathbf{e}_+$-elliptic polarization of undoped type-I [panel (a)] and type-II WSM [panel (b)], 
	for different values of $\beta$ ($= 0, 0.1, 0.2, 0.4$ and $0.6$).  The curves have been displaced up for increasing value of $\beta$. 
	In the case $\beta = 0$ one observes the usual Landau level spectrum with $n \rightarrow (n\pm1)$ transitions 
	while for higher values of the in-plane tilt, more transitions are allowed. Insets are a reminder of the LL structure and the arrows indicate inter-LL
	transitions visible in the optical conductivity.}
	\label{peaks}
\end{figure}

We show however that distinctive signatures of type-I and type-II WSM 
are clearly accessible in
magnetooptical (reflectivity) experiments \cite{mikane} in the magnetic regime,
with no electric field involved.
Within linear-response theory, the reflection of light at a polarization $\mathbf{l}$ is proportional to the real part of the optical conductivity 
\begin{eqnarray}\label{eq:reop}
 \mathrm{Re}~\sigma_{ll}(\omega) &= &\frac{\sigma_0}{2\pi l_B^2\omega} \sum_{\substack{j,j'}}|\mathbf{l}\cdot \mathbf{V}_{j,j'}|^2 \nonumber
    \\
    &&\times \left[f(E_{j}) - f(E_{j'})\right]\delta(\omega - \omega_{j j'})
\end{eqnarray}
where the subscripts $j,j'$ denote the quantum numbers $(\lambda,n,k_z)$, $\sigma_0=e^2/2\pi$ is the quantum of conductance, and $\omega_{jj'}=E_j - E_{j'}$ is
the energy difference between the final and initial states. Furthermore, $f(E_j)$ is the Fermi-Dirac distribution for a Fermi level that we choose at zero energy
(at the position of the Weyl point) and the matrix element is given by $\mathbf{V}_{j,j'} = \langle \Psi_{j} | \nabla_{k} H_B | \Psi_{j'} \rangle$. This formula
is integrated analytically for two elliptic polarizations (see Supplementary Material). 
One of the most salient features of the matrix elements in Eq. (\ref{eq:reop}) is related to the nonorthogonality of the spinor components as a consequence of
the relativistic boost \cite{judit}, i.e. when $\beta\neq 1$. Indeed, the harmonic-oscillator wave functions acquire an energy-dependent shift in their position
due to the Lorentz transformation, which thus yields energy-dependent overlap functions (see Supplementary Material). 
As a consequence, the usual dipolar selection $n\rightarrow (n\pm 1)$
are violated, and as a function of $\beta$ additional interband peaks arise in the optical conductivity at energies
\begin{equation}
     \omega_{mn} = (1-\beta^2)^{3/4} \sqrt{2eB}v_{\perp}\left( \sqrt{m} + \sqrt{n} \right).
\end{equation}
This is shown in Fig. \ref{peaks} for a type-I [panel (a)] and
a type-II WSM [panel (b)], where we depict the optical conductivity as a function of frequency for several values of $\beta$. For $\beta=0$, one retrieves the
typical spectrum of a WSM discussed in the literature \cite{Ashby,tabert}, where the peaks have the usual $1/\sqrt{\omega-\omega_{mn}}$ 
divergence due to the 1D character of the 
LL bands as a function of $k_z$. These are also visible at $\beta\neq 0$ and are the main difference with respect to the 2D case discussed in Ref. \cite{judit}. 
Furthermore, the optical conductivity of an undoped 
type-II WSM shows additional peaks at low energies that are due to allowed transitions between LLs of the same original band. Indeed, LLs 
corresponding to the family $\lambda=+$, which are always at positive energies in a type-I WSM, have parts of bands situated at negative energies, whereas those
of with $\lambda=-$ become positive for certain values of $k_z$. Therefore, transitions between LLs of the same family [red arrows in the inset of panel (a)]
are no longer blocked by the Pauli principle and contribute to the optical conductivity up to the energy
\begin{equation}\label{eq:edge}
	\omega_{l} =  (1-\beta^2)^{3/4}\sqrt{2eB} v_{\perp} \frac{\sqrt{2} |t_z| - \sqrt{t_z^{2} + 1}}{\sqrt{t_z^{2} -1}}.
\end{equation}
This yields the unusually large optical conductivity at low frequency 
and reflects the large density of states of a type-II WSM at zero $B$-field, in comparison with a type-I WSM. The effect is different from that previously 
described in Ref. \cite{Ashby} for a type-I WSM where low-frequency peaks in the optical conductivity are obtained upon doping, whereas here the system is undoped.

In conclusion, we have shown that 3D WSM show a rich variety of phases in the presence of a magnetic field that unveil the underlying relativistic symmetry. 
In contrast to 2D systems, the magnetic and the electric regimes, with and without LL quantization, respectively, do no longer
coincide with the type of the WSM in 3D. Indeed, a type-II WSM can show
LL quantization as long as the angle between the tilt direction and the magnetic field satisfies the condition (\ref{eq:critang})--in this case, the type-II character
is visible in the 1D Landau bands as a function of the wave vector $k_z$ in the direction of the $B$-field. 
Most saliently, we have shown that the relativistic parameter
$\beta$ has a clear fingerprint in the optical conductivity, which can be probed in magnetooptical measurements, discerning different types of WSM. 
We would finally emphasize that our results are readily generalized to the case of a pair (or more) of Weyl points in a more realistic band structure. If the two 
cones are related by a discrete symmetry (e.g. time reversal), the tilts of the two cones are in opposite directions. In this case, one obtains two copies of the 
same LL spectrum, since the parameter $\beta$ depends only on the modulus of the tilt vector ${\bf t}_{\perp}$. For more pairs of Weyl points, the tilt vectors are 
no longer necessarily of the same modulus, and the condition (\ref{eq:critang})
for the transition between magnetic and electric regime is then not unique for all points.
During the writing of the present paper, we became aware
of two other preprints on magnetic-field properties of type-II WSM. The findings of Ref. \cite{yuan} are similar to ours but 
the authors consider only the optical conductivity
for $\beta=0$, i.e. when the tilt is in the same direction as the $B$-field, in which case one has no violation of the dipolar selection rule $n\rightarrow (n\pm 1)$ 
and no renormalization of the LL spacing. The authors of Ref. \cite{bergholz} study the chiral anomaly furthermore in a lattice tight-binding model and extract
numerically the optical conductivity of a type-II WSM in the magnetic regime. 

We acknowledge fruitful discussions with Milan Orlita and Marek Potemski.

\pagebreak

\setcounter{equation}{0}
\setcounter{figure}{0}

\renewcommand{\theequation}{S\arabic{equation}}
\renewcommand{\thefigure}{S\arabic{figure}}

\begin{widetext}

\section*{Supplementary Material}

\section{Minimal model for tilted Weyl semimetal}
We consider the model of a tilted Weyl Hamiltonian with an anisotropic Fermi velocity
\begin{align}
	H = \omegab_0\cdot \mathbf{k} \bone + \sum_{\mu = 1}^3 v_{\mu} k_{\mu} \hat{\sigma}_{\mu},
	\label{wsm0}
\end{align}
where $\hat{\sigma}_{\mu}$ ($\mu = 1,2,3$) are the three Pauli matrices, the $v_{\mu}$ are the anisotropic Fermi velocities and $\mathbf{\omega_0}$ is the tilt velocity. Upon rescaling of the momenta
\begin{align}
	q_{\mu} = \frac{v_{\mu}}{v_{\perp}} k_{\mu}~\textrm{for } \mu \in \{1,2\},
\end{align} 
the Hamiltonian reads
\begin{align}
	H = \omegab_1\cdot \mathbf{q} \bone + v_{\perp}(q_x\hat{\sigma}_x + q_y\hat{\sigma}_y) + v_zk_z\hat{\sigma}_z,
\end{align}
where the anisotropy has been absorbed in the tilt parameters, $\omega_{1,\mu} = v_{\perp} \omega_{0,\mu}/v_{\mu}$ for $\mu \in \{1,2\}$ and $\omega_{1,3} = \omega_{0,z}$. 
The Hamiltonian can still be simplified by the following rotation
\begin{align}
	\left\{
	\begin{array}{l}
	H_B = e^{i\frac{\phi}{2} \hat{\sigma}_z} H e^{-i\frac{\phi}{2} \hat{\sigma}_z} \\
	|\Psi\rangle = e^{i\frac{\phi}{2} \hat{\sigma}_z} |\Psi_{\phi}\rangle\\
	q_x' = \cos(\phi) q_x + \sin(\phi) q_y~~~,\\
	q_y' = \cos(\phi) q_y - \sin(\phi) q_x\\
	q_z' = k_z
	\end{array}
	\right.
	\label{rot}
\end{align}
with $\tan(\phi) = \omega_{0,y}/\omega_{0,x}$, one finds the minimal form of the Hamiltonian $H_B$, Eq. (3) %(\ref{eq:ham2}) 
of the main text. 
We emphasize that this rotation, which does not depend on the wave vectors and keeps $k_z$ invariant, leaves the commutation rules between the $x$- and $y$-components
of the wave vector unchanged, even in the presence of a magnetic field. We consider this form of the Hamiltonian in the main text.

\section{Eigenstates in the magnetic regime}
\subsubsection{Magnetic regime} 
In the magnetic regime introduced in the main text (5), %(\ref{mal}), 
the momenta fulfil the 
following commutation rules [we consider $\mathrm{sign}(v_xv_yB) = 1$]
\begin{align}
	\begin{array}{ccc}
		[\tilde{q}_x,\tilde{q}_y] = - \frac{i}{\gamma l_B^2},	&	[\tilde{q}_x,\tilde{k}_z] = 0,	&	[\tilde{q}_y,\tilde{k}_z] = 0~,
	\end{array}
\end{align}
which indicate that the magnetic field is lowered by $B \rightarrow B_{\mathrm{eff.}} = \sqrt{1 - \beta^2} B$.  We use the following ladder operators 
\begin{align}\label{eq:a}
	\left\{
	\begin{array}{l}
	\hat{a}(E,k_z)  = \sqrt{\frac{\gamma}{2}} l_B  \left[ \tilde{q}_x(E,k_z) - i \tilde{q}_y \right], \\~~\\
	\hat{a}^{\dagger}(E,k_z)  = \sqrt{\frac{\gamma}{2}} l_B  \left[ \tilde{q}_x(E,k_z) + i \tilde{q}_y \right] ,
	\end{array}\right.
\end{align}
and solve the Landau levels equation (5) %(\ref{mal}) 
for $n > 0$,
\begin{align}\label{eq:sn}
	\left\{
	\begin{array}{l}
	E_{n,\pm}(k_z) = \omega_{0,z} k_z \pm \frac{1}{\gamma} \sqrt{v_z^2 k_z^2 + \frac{ 2 eBv_{\perp}^2}{\gamma}n }\\
	|\tilde{\Psi}_{n,\pm}\rangle = \frac{1}{\sqrt{2}}\left[
		\begin{array}{c}
			\left( 1 \pm \frac{v_z k_z}{\Delta E_n}\right)^{\frac{1}{2}} |n-1\rangle\\
			\pm \left( 1 \mp \frac{v_z k_z}{\Delta E_n}\right)^{\frac{1}{2}}|n\rangle
		\end{array}
	\right],
	\end{array}
	\right.
\end{align}
where $\Delta E_n = \sqrt{ v_z^2 k_z^2 + \frac{ 2 eB v_{\perp}^2 }{\gamma}n }$, and for $n = 0$
\begin{align}\label{eq:s0}
	\left\{
	\begin{array}{l}
	E_{0}(k_z) = \left(\omega_{0,z} -  \frac{v_{z}}{\gamma}\right) k_z\\
	|\tilde{\Psi}_{0}\rangle = \left[
		\begin{array}{c}
			0\\
			| 0 \rangle
		\end{array}
	\right].
	\end{array}
	\right.
\end{align}
The states in original basis are obtained by $|\Psi\rangle = (1/\Nmath)e^{\frac{\theta}{2}\hat{\sigma}_x}|\tilde{\Psi}\rangle$, \emph{i.e.} 
the Lorentz boost mixes the components.

\subsubsection{Number states } In Eqs. (\ref{eq:sn}) and (\ref{eq:s0}), the $|n\rangle$ states correspond to the eigenstates of the number operator, 
$\hat{n}(E,k_z) = \hat{a}^{\dagger}(E,k_z) \hat{a}(E,k_z)$, that depends explicitly on energy $E$ and momentum $k_z$. In the following we 
explicitly write the number states with their $E$- and $k_z$-dependence.

Two number states $|n_1,E_1,k_{z,1}\rangle$ and $|n_2,E_2,k_{z,2}\rangle$ are in general not orthogonal.  Indeed, the difference between two ladder operators $\hat{a}(E_1,k_{z,1})$ and $\hat{a}(E_2,k_{z,2})$ defined in (\ref{eq:a}) of different energies, $E_1$ and $E_2$, and momenta, $k_{z,1}$ and $k_{z,2}$, is a scalar
\begin{align}\label{eq:disp}
	\hat{a}(E_1, k_{z,1}) - \hat{a}(E_2, k_{z,2}) &= \alpha_{1,2},
\end{align}
where
\begin{align}
	&\alpha_{1,2} =  \alpha(E_1 - E_2, k_{z,1} - k_{z,2}),\\
	&\alpha(E, k_z) = \frac{\beta}{\left( 1 - \beta^2 \right)^{3/4}} \frac{\omega_{0,z} k_z - E}{\sqrt{2eB} v_F}.
\end{align}
The scalar shift $\alpha_{1,2}$ signifies that the cyclotron orbits at different energies and momenta are displaced from each other \cite{coher}.
For this reason the $|0,E,k_z\rangle$-states (\emph{i.e.} such that $\hat{a}(E,k_{z})|0,E,k_z\rangle = 0$) 
depend on energy and momentum. If one applies (\ref{eq:disp}) to $|0,E_2,k_{z,2}\rangle$ one finds the following coherent-state equation
\begin{align}\label{eq:coher}
	&\hat{a}(E_1,k_{z,1}) | 0,E_2,k_{z,2}\rangle  = \alpha_{1,2} |0,E_2,k_{z,2} \rangle ,	
\end{align}
because $\hat{a}(E_2,k_{z,2})|0,E_2,k_{z,2}\rangle = 0$ by construction. One then deduces \cite{coher} the following relation between two number 
basis of different $(E,k_z)$
\begin{align}
	&| 0, E_2, k_{z,2}\rangle = e^{-\frac{|\alpha_{1,2}|^2}{2}} \sum_{p = 0}^{\infty} \frac{\alpha_{1,2}^p}{\sqrt{p!}}| p, E_1, k_{z,1} \rangle.
\end{align}
This coherent-state property has consequences on the scalar product of $|n, E, k_{z} \rangle$-state which is by construction
\begin{align}
	&\langle n_{2}, E_2, k_{z,2} | n_1, E_1, k_{z,1} \rangle\\ &= \frac{1}{\sqrt{n_2!}} \langle 0, E_2, k_{z,2} | \hat{a}(E_2,k_{z,2})^{n_2}| n_1, E_1, k_{z,1} \rangle.\nonumber
\end{align}
We replace $\hat{a}(E_2,k_{z,2})$ by its expression in Eq. (\ref{eq:disp}) and use the binomial theorem $(a + b)^{n_2} = \sum_{k = 0}^{n_2} {{n_2} \choose{k}}a^k b^{n_2-k}$. Also we replace $\langle 0, E_2, k_{z,2} |$ by its expression as a coherent state of $\hat{a}(E_1,k_{z,1})$ as in (\ref{eq:coher}). The terms that appear in the sum of the binomial theorem are of the form
\begin{align}
	\langle \alpha_{1,2} | &\hat{a}(E_1,k_{z,1})^{n_2 - k} | n_1, E_1, k_{z,1} \rangle\nonumber\\
	 &= \sqrt{\frac{n_1!}{(k + n_1 - n_2)!}} \langle \alpha_{12}| k + n_1 - n_2, E_1, k_{z,1} \rangle\nonumber\\
	 &= \frac{\sqrt{n_1!}\alpha_{12}^{*k + n_1 - n_2}}{(k + n_1 - n_2)!}e^{-\frac{|\alpha_{12}|^2}{2}}.
\end{align}
If one introduces this term back in the sum, one recognizes the generalized Laguerre polynomial $L^{(\alpha)}_{n}(x) = \sum_{k=0}^{n} { n+\alpha \choose{n-k} }\frac{(-x)^k}{k!}$ and one finds
\begin{align}\label{eq:scalar}
	\langle n_{2}, E_2, k_{z,2} &| n_1, E_1, k_{z,1} \rangle \\&= \sqrt{\frac{n_2!}{n_1!}}\alpha_{1,2}^{*n_1 - n_2} L_{n_2}^{(n_1 - n_2)}(|\alpha_{1,2}|^2) e^{-\frac{|\alpha_{1,2}|^2}{2}}\nonumber.
\end{align}

\section{Magneto-optics}
\label{optc}
The nonorthogonality (\ref{eq:scalar}) of the spinor components has drastic consequences 
on the Landau levels spectroscopy of tilted Dirac cones, as we expose in detail here. 

\subsubsection{Light-matter interaction } 
The electron-light coupling is of the form
\begin{align}
	\hat{H}_{\mathrm{int}} &= e\mathcal{A}\cdot \mathbf{\nabla_{k}}\hat{H},
\end{align}
where $\mathcal{A}$ is the vector potential corresponding to the oscillating electric field $\mathcal{E}(\omega)$. For convenience we study this coupling 
term in the orientation defined by Eq. (\ref{rot}) and in the basis defined by the Lorentz boost in Eq. (5) % (\ref{mal}) 
of the main text
\begin{align}
	\hat{H}_{B,\mathrm{int}} &=  \mathcal{A} \cdot e^{\frac{\theta}{2}\hat{\sigma}_x} \left( \mathbf{\nabla_{k}}\hat{H}_{B}\right) e^{\frac{\theta}{2}\hat{\sigma}_x}\\
	&= \mathcal{A} \cdot  \mathbf{\nabla_{k}}\left( e^{\frac{\theta}{2}\hat{\sigma}_x} \hat{H}_{B} e^{\frac{\theta}{2}\hat{\sigma}_x}  \right)\\
	&= \left( \mathcal{A}\cdot \mathbf{e}_{+} \right) v_{e} \hat{\sigma}_+ + \left( \mathcal{A}\cdot \mathbf{e}_{-} \right) v_{e} \hat{\sigma}_- + \left( \mathcal{A}\cdot \mathbf{e}_z \right) v_z \hat{\sigma}_z,
\end{align}
where we use the fact that the boost $\beta$ is independent of the momenta. We introduced two 
convenient -- but not orthogonal, \emph{i.e.} $\mathbf{e}_{+}\cdot \mathbf{e}_{-} \neq 0$ -- elliptical polarization modes $\mathbf{e}_{\pm}$ defined as
\begin{align}
	&\mathbf{e}_{\pm} = \frac{1}{v_{e}}
	\left[
		\begin{array}{c}
			v_x/{\gamma}\\
			\mp i v_y \\
			0
		\end{array}
	\right],
	\label{epol}\\
	&\mathbf{e}_{z} = \left[
		\begin{array}{c}
		0\\
		0\\
		1
		\end{array}
	\right],
\end{align}
where $v_{e} = \sqrt{v_x^2/\gamma^2 + v_y^2}$ and the Cartesian axis are in the directions defined in Eq. (\ref{rot}). We also defined $\hat{\sigma}_{\pm}$ as
\begin{align}
	\hat{\sigma}_{\pm} = \frac{1}{2}\left( \hat{\sigma}_x \pm i\hat{\sigma}_y \right).
\end{align}

\subsubsection{Optical conductivity}
We consider optical measurements in the Faraday geometry where the light polarisation $\mathbf{l}$ (\emph{i.e.} $\mathcal{A} \sim \mathbf{l}$) satisfies 
$\mathbf{l}\cdot \mathbf{e}_z = 0$. The light reflection at polarization $\mathbf{l}$ is proportional to the real part of the optical conductivity 
$\sigma_{ll}(\omega)$, defined as $\mathbf{j}\cdot \mathbf{l} = \sigma_{ll}(\omega) \mathcal{E}\cdot\mathbf{l}$. We use the expression (11) % (\ref{eq:reop}) 
introduced 
in the main text obtained from linear response theory and consider the chemical potential $\mu = 0$, zero temperature, and $\omega > 0$.  In the case of a 
type-I Weyl semimetal, the band occupation term is non zero for $\lambda = - \rightarrow +$ transitions only. Once integrated, one finds the following expression 
for the optical conductivity
%\begin{widetext}
\begin{align}
	\mathrm{Re}\left[\sigma^I_{ll}(\mathbf{\omega})\right] = \frac{v_{e}^2\sigma_0 n_B }{2\pi\omega v_z} &\left(  |\mathbf{l}\cdot \mathbf{e}_{+}|^2\sum_{n \in \mathbb{N}} \Theta\left[ \omega - (1 - \beta^2)^{3/4}v_{\perp}\sqrt{2eBn}\right]R_{0}^{n-1}(\omega)^2\nonumber\right.\\ &\left.+ \sum_{\substack{n,m \in \mathbf{N}^*}}\frac{\Theta\left[ \omega - (1 - \beta^2)^{3/4}v_{\perp}\sqrt{2eB} \left( \sqrt{m} + \sqrt{n} \right)\right]}{\sqrt{\left[ (\gamma  \omega)^2 - \frac{2eBv_{\perp}^2}{\gamma}(n+m)\right]^2+ \left(\frac{2eBv_{\perp}^2}{\gamma}\right)^2\left[(n-m)^2 -(n+m)^2\right]}}\right.\\
	& \left\{ 2\left[ (\gamma  \omega)^2 - \frac{2eBv_{\perp}^2}{\gamma}(n+m) \right]\left[|\mathbf{l}\cdot \mathbf{e}_{+}|^2 R_{m}^{n-1}( \omega)^2+ |\mathbf{l}\cdot \mathbf{e}_{-}|^2R_{m-1}^{n}( \omega)^2\right] \right. \nonumber\\
	&\nonumber\left. \left. -  \frac{4eBv_{\perp}^2}{\gamma} \sqrt{nm} \textrm{ Re}\left[ \left( \mathbf{l}\cdot \mathbf{e}_{+} \right)^*\left( \mathbf{l}\cdot \mathbf{e}_{-} \right)\right]  R_{m-1}^{n}( \omega)R_{m}^{n-1}( \omega)\right\}\right),
\end{align}
%\end{widetext}
where $\Theta(x)$ is the Heaviside function, $n_B=1/2\pi l_B^2$, and $R_{p}^{s}( \omega)$ is the overlap function introduced in Eq. (\ref{eq:scalar}) at energy $\omega$,
\begin{align}\label{eq:select}
	R_{p}^{s}( \omega) &= \langle s, E_{n,+}, k_z| p, E_{m,-}, k_z \rangle\\
	&=\sqrt{\frac{s!}{p!}}\alpha_{\omega}^{p - s} L_{s}^{(p - s)}(|\alpha_{\omega}|^2) e^{-\frac{|\alpha_{\omega}|^2}{2}},\\
	\alpha_{\omega} &= -\frac{\beta}{\left( 1 - \beta^2 \right)^{3/4}} \frac{ \omega}{\sqrt{2eB} v_{\perp}}.
\end{align}
The expression in the case of a type-II Weyl semimetal also includes $\lambda = \pm \rightarrow \pm $ transitions the contributions of which are similar to the previous $-\rightarrow +$ transitions. We do not show the corresponding expression here but one has to introduce the band edge energy scales
\begin{align}\label{eq:fedge}
	 &\omega^{(\mathrm{edge})}_{nm, \lambda\lambda'} =\\ &(1-\beta^2)^{3/4}\sqrt{2eB} v_{\perp} \left[ \frac{\lambda |t_z| \sqrt{n} - \lambda' \sqrt{ n + \left[t_z^{2} - 1\right] m}}{\sqrt{t_z^{2} - 1}} \right].\nonumber
\end{align}
This expression of the band edge is the reason for the low-frequency intraband peak width (13) % (\ref{eq:edge}) 
introduced in the main text. One other property of the band edge 
is that it is given by $\omega^{(\mathrm{edge})}_{m0,\lambda s}$ where $s = \mathrm{sign}(v_xv_yv_z\omega_{0z}B)$ for the transitions involving the $n = 0$ state. This is not discussed in the main text but it can shift the edges depending on the direction of the magnetic field.

\end{widetext}

\end{document}